\documentclass[english,aps,prl,twocolumn,showpacs,superscriptaddress]{revtex4-1}
\usepackage[T1]{fontenc}
\usepackage[latin9]{inputenc}
\usepackage{xcolor}
\usepackage{amsmath}
\usepackage{graphicx}
\usepackage{ulem}
\usepackage{babel}

\begin{document}

\title{Quantum Routing of Single Photons with Cyclic Three-Level System}

\author{Lan \surname{Zhou}}
\affiliation{Key Laboratory of Low-Dimensional Quantum Structures and Quantum
Control of Ministry of Education, and Department of Physics, Hunan
Normal University, Changsha 410081, China}
\author{Li-Ping Yang}
\affiliation{State Key Laboratory of Theoretical Physics, Institute of Theoretical
Physics, University of Chinese Academy of Science, Beijing 100190,
China }
\author{Yong Li}
\affiliation{Beijing Computational Science Research Center, Beijing 100084, China}
\author{C. P. Sun}
\email{cpsun@csrc.ac.cn}
\homepage{http://www.csrc.ac.cn/~suncp}
\affiliation{Beijing Computational Science Research Center, Beijing 100084, China}

\begin{abstract}
We propose an experimentally accessible single-photon routing scheme
using a $\bigtriangleup$-type three-level atom embedded in quantum
multi-channels composed of coupled-resonator waveguides. Via the on-demand
classical field being applied to the atom, the router can extract a single
photon from the incident channel, and then redirect it into another.
The efficient function of the perfect reflection of the single-photon signal
in the incident channel is rooted in the coherent resonance and the
existence of photonic bound states.
\end{abstract}

\pacs{03.65.Nk, 03.67.Lx, 42.50.Ex, 78.67.-n}
\maketitle

Scalable quantum information processing in quantum computation and
communication is essentially based on a quantum network. A key element
inside is the quantum node, which coherently connects different quantum
channels. It is named as quantum router for controlling the path of the
quantum signal with fixed Internet Protocol (IP) addresses, or named as
quantum switch without fixed IP addresses.

Recently, many theoretical proposals and experimental demonstrations of
quantum router have been carried out in various systems, i.e., cavity QED
system~\cite{PRL09_Kimble}, circuit QED system~\cite{CrQED}, optomechanical
system~\cite{optom}, and even a pure linear optical system~\cite%
{opticzei,opticDuan}. The essence lying in core is the realization of the
coupling between a two- (or few-) level system and quantum channels \cite%
{fan1ph,fan2ph,sun2l,Longo,sun3l}. We notice that, except the experiment
in Ref.~\cite{opticDuan} implemented with linear optical devices, the
quantum router demonstrated in most experiments and theoretical proposals
has only one output terminal. Thus the ideal quantum router with
multi-access channels deserves more exploration.

In this letter, we theoretically propose a scheme for quantum routing of
single photons with two output channels, which are composed of two
coupled-resonator waveguides (CRWs). While the quantum node is realized by a
three-level system with three transitions forming a cyclic ($\bigtriangleup$%
-type) structure~\cite{cyclicYL,PRL01_Chiral_molecule,cycleAL,LYPRL99}. To
locate the different IP addresses, two different transitions of the $%
\bigtriangleup$-atom are coupled to the photonic modes of the two
channels respectively and the other is used to connect the two channels with
a classical field. We study the single-photon scattering in this proposed
hybrid system. It is shown that the quantum node indeed works as a
multi-channel quantum router since the classical field can redirect single
photons into different channels. The total reflection are guaranteed by the
existence of quasi-bound states due to the coupling of a discrete energy
level and a continuum. Actually, there have been numerous theoretical
studies~\cite{BS1,BS2,BS3,BS4,Dazhi,JMZhang} focusing on the von Neumann-Wigner
conjecture~\cite{BIC}: whether or not there exist (quasi-) bound states when
discrete energy levels are coupled to a continuum. Now, our hybrid system
provides a platform to probe this kind of bound states.

\begin{figure}[tbp]
\includegraphics[bb=0bp -1bp 688bp 400bp,width=8cm]{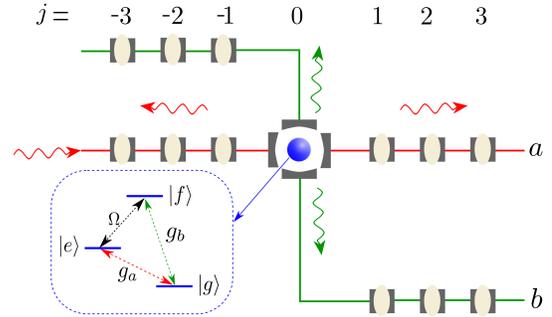}
\caption{(Color online) Schematic of routing single photons in two channels
made of two CRWs. The three-level atom characterized by $\left|g\right%
\rangle $, $\left|e\right\rangle $, and $\left|f\right\rangle $ is placed at
the cross point $j=0$. CRW-$a$ (-$b$) couples to the atom through the
transition $\left|g\right\rangle \leftrightarrow\left|e\right\rangle $ ($%
\left|g\right\rangle \leftrightarrow\left|f\right\rangle $) with strength $%
g_{a}$ ($g_{b}$) and a classical field $\Omega$ is applied to resonantly
drive $\left|e\right\rangle \leftrightarrow\left|f\right\rangle $
transition. An incoming wave from the left side of CRW-$a$ will be
reflected, transmitted, or transferred to CRW-$b$.}
\label{fig:1}
\end{figure}

\textit{Quantum node with $\bigtriangleup$-type atom}.---The
considered system (see Fig.~\ref{fig:1}) consists of two one-dimensional
(1D) CRWs channels whose cavity modes are described by the creation
operators $a_{j}^{\dagger}$ and $b_{j}^{\dagger}$, respectively, and a $%
\bigtriangleup$-type three-level atom characterized by a ground-state $%
\left|g\right\rangle $ and two excited states $\left|e\right\rangle $ and $%
\left|f\right\rangle $. The atom at $j=0$ resonators connects two CRWs since
the cavity modes $a_{0}$ and $b_{0}$ enable the dipole-allowed transitions $%
\left|g\right\rangle \leftrightarrow\left|e\right\rangle $ and $%
\left|g\right\rangle \leftrightarrow\left|f\right\rangle $ with coupling
constants $g_{a}$ and $g_{b}$, respectively. A classical field with
frequency $\nu=\omega_{e}-\omega_{f}$ resonantly drives the transition $%
\left|e\right\rangle \leftrightarrow\left|f\right\rangle $ with Rabi
frequency $\Omega$. Usually, such cyclic systems are forbidden for nature
atoms, but can exist in symmetry-broken systems~\cite{cyclicYL}, e.g.,
chiral molecules~\cite{PRL01_Chiral_molecule} and artificial symmetry-broken
atoms~\cite{cycleAL,LYPRL99}.

In the rotating frame with respect to
\begin{equation*}
H_{0}=\omega_{e}(\sum_{j}a_{j}^{\dagger}a_{j}+\left|e\right\rangle
\left\langle
e\right|)+\omega_{f}(\sum_{j}b_{j}^{\dagger}b_{j}+\left|f\right\rangle
\left\langle f\right|),
\end{equation*}
the Hamiltonian of two CRWs is described by a typical tight-binding bosonic
model,
\begin{equation}
H_{c}=\sum_{d=a,b}\sum_{j}\left[\Delta_{d}d_{j}^{\dagger}d_{j}-%
\xi_{d}(d_{j+1}^{\dagger}d_{j}+\mathrm{h.c.})\right],
\end{equation}
where $\xi_{d}$ are the homogeneous inter-cavity coupling constants and $%
\Delta_{a(b)}=\omega_{a(b)}-\omega_{e(f)}$ are cavity-atom detunings.
Hereafter $d$ stands for $\{a,b\}$. Under the rotating-wave approximation,
the interaction between the atom and two CRWs is written as
\begin{equation}
H_{\mathrm{int}}=g_{a}\left|e\right\rangle \left\langle
g\right|a_{0}+g_{b}\left|f\right\rangle \left\langle
g\right|b_{0}+\Omega\left|e\right\rangle \left\langle f\right|+\mathrm{H.c.}
\end{equation}

The two photonic channels are illustrated by the red and green lines in Fig.~%
\ref{fig:1}. Obviously, in the absence of the classic field, single photons
incident from the red channel get reflected or transmitted only within the
red one. Photons incident from the red channel can be switched into the
other (green) channel only when $\Omega\neq0$.

\textit{Coherent scattering of single photons}.---Fourier transformation $%
d_{k}=(1/\sqrt{N})\sum_{j}d_{j}\exp(ikj)$ shows that each bare CRW supports
plane waves with dispersion relation $E_{k}^{(d)}=\Delta_{d}-2\xi_{d}\cos
k_{d}$ ($k_{d}\in\left[0,2\pi\right])$, i.e., each CRW possesses an energy
band with the bandwidth $4\xi_{d}$. The detunings $\Delta_{d}$ and the
inter-cavity coupling constants determine whether the two bands overlap or
not. The classical field mixes the two bands.

Inside the photonic band, the incident photon with energy $E$ will be
elastically scattered. The single-excitation eigenstate is supposed to be
\begin{equation}
\left|E\right\rangle =U_{e}\left|e0\right\rangle +U_{f}\left|f0\right\rangle
+\sum_{j}(A_{j}a_{j}^{\dagger}+B_{j}b_{j}^{\dagger})\left|g0\right\rangle ,
\end{equation}
where $\left|0\right\rangle $ is the vacuum state of the CRWs and $U_{e}$, $%
U_{f}$, $A_{j}$, and $B_{j}$ are the corresponding amplitudes. The motion of
single photons are governed by the discrete scattering equations
\begin{eqnarray}
ED_{j} & = & \Delta_{d}D_{j}-\xi_{d}\left(D_{j+1}+D_{j-1}\right)  \notag \\
& & +\delta_{j0}\left[V_{d}\left(E\right)D_{j}+G\left(E\right)\bar{D}_{j}%
\right],  \label{eq:scat}
\end{eqnarray}
which is obtained from the Schr\"{o}dinger equation with Hamiltonian
$H=H_{c}+H_{\mathrm{int}}$ by reducing the atomic amplitudes. Here, $D=\{A,\ B\}$
and the over bar designates the element other than $D$ in the set $\left\{ A,B\right\}$.
For convenience, we have introduced the energy-dependent delta-like potentials with strength
$V_{d}\left(E\right)=Eg_{d}^{2}/\left(E^{2}-\Omega^{2}\right)$ and the effective
dispersive coupling strength $G\left(E\right)=\Omega g_{a}g_{b}/\left(E^{2}-\Omega^{2}\right)$
between the resonator modes $a_{0}$ and $b_{0}$. The coupling $G$ leads to
the channel-switching. At $\left|E\right|=\Omega$, infinite delta potentials
are formed at $j=0$ in both CRWs. It seems that the delta potential would
prevent the propagation of single photons. However, the effective coupling
strength $G$ also becomes infinite at $\left|E\right|=\Omega$, which might
transfer the photon from one CRW to the other.

A wave with energy $E$ incident from the left side of one CRW (says CRW-$a$)
will result in reflected, transmitted, and transfer waves with the same
energy. The wave functions in the asymptotic regions are given respectively
by
\begin{equation}
A(j)=%
\begin{cases}
e^{ik_{a}j}+r^{a}e^{-ik_{a}j}, & j<0 \\
t^{a}e^{ik_{a}j}, & j>0%
\end{cases}
\label{eq:ansatz-ca}
\end{equation}
and
\begin{equation}
B(j)=%
\begin{cases}
t_{l}^{b}e^{-ik_{b}j}, & j<0 \\
t_{r}^{b}e^{ik_{b}j}, & j>0%
\end{cases}
\label{eq:ansatz-cb}
\end{equation}
where $t^{a}$ ($r^{a}$) is the transmitted (reflected) amplitude and $%
t_{l}^{b}$ ($t_{r}^{b}$) is the forward (backward) transfer amplitude.
Applying solutions~(\ref{eq:ansatz-ca}) and~(\ref{eq:ansatz-cb}) to the
discrete scattering equations~(\ref{eq:scat}), we obtain the scattering
amplitudes,
\begin{align}
\!\!\!\!\! t^{a}(E) & =\frac{2i\xi_{a}\sin k_{a}\left[2i\xi_{b}\sin
k_{b}-V_{b}\left(E\right)\right]}{\prod_{d=a,b}\left[2i\xi_{d}\sin
k_{d}-V_{d}\left(E\right)\right]-G^{2}\left(E\right)},  \label{eq:tal-to-r}
\\
\!\!\!\! t^{b}(E) & =\frac{2i\xi_{a}\sin k_{a}G\left(E\right)}{\prod_{d=a,b}%
\left[2i\xi_{d}\sin k_{d}-V_{d}\left(E\right)\right]-G^{2}\left(E\right)},
\label{eq:tbl-to-r}
\end{align}
with continuity conditions $t_{r}^{b}=t_{l}^{b}\equiv t^{b}$ and $%
t^{a}=r^{a}+1$. Here, wave numbers $k_a$ and $k_b$ satisfy
$E=\Delta_a-2\xi_a\cos k_a=\Delta_b-2\xi_a\cos k_b$. From Eqs.~(\ref{eq:tal-to-r})
and~(\ref{eq:tbl-to-r}), we find that turning on the classical field
makes the incoming wave in CRW-$a$ transfer to CRW-$b$. It is the classical
field that implements the photon-redirection function of the quantum router.

\begin{figure}[tbp]
\includegraphics[height=8cm]{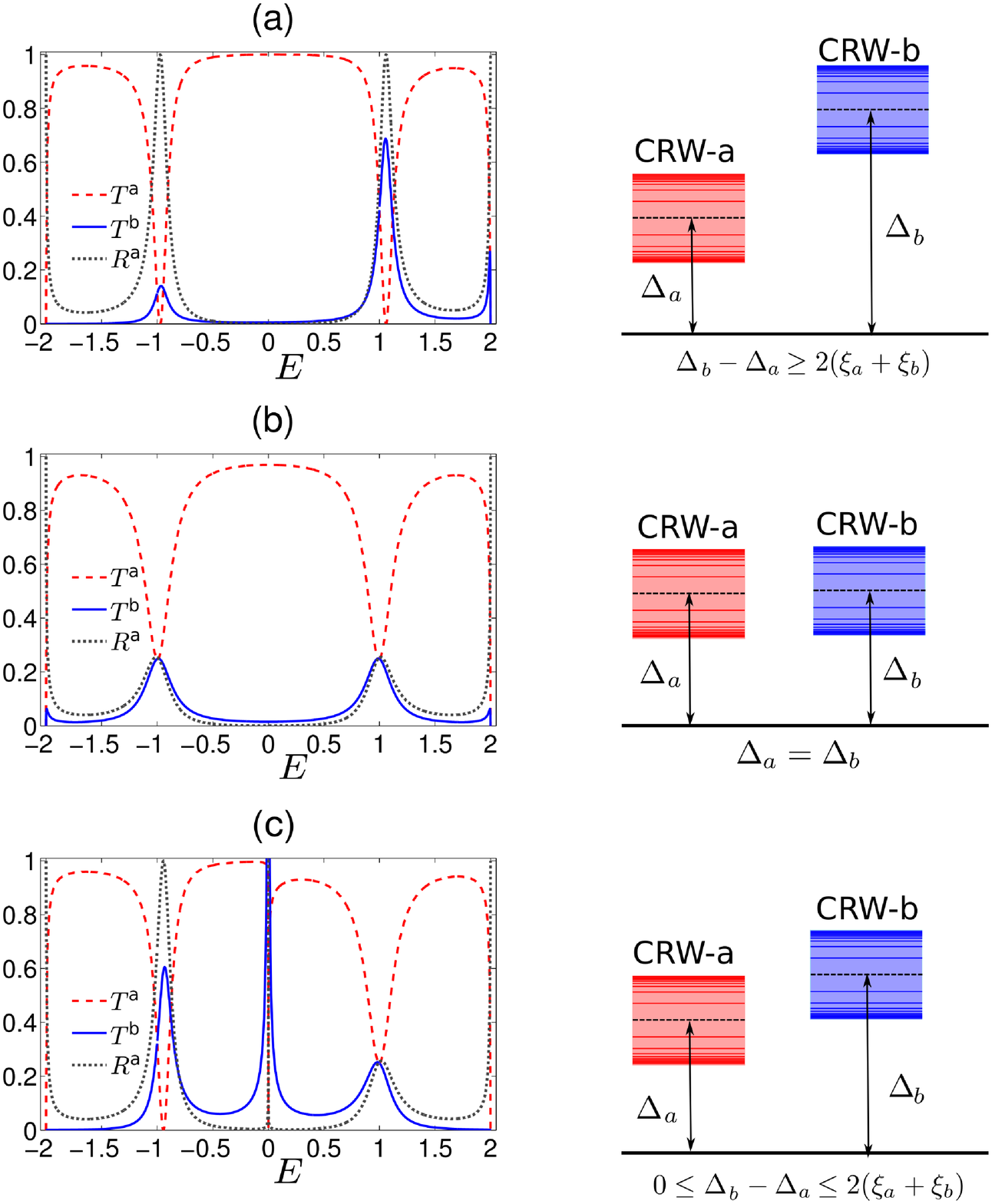}
\caption{(Color online) The scattering process described by transmittance $T^{a}(E)$
(dashed red line), reflectance $R^{a}(E)$ (dotted gray line) and transfer rate
$T^{b}(E)$ (solid blue line) with different configuration
of the two bands of the vacant CRWs. (a) $\Delta_{b}=4.5$, thus
$\Delta_{b}-\Delta_{a}>2(\protect\xi_{a}+\protect\xi_{b})$; (b) $\Delta_{b}=0$;
(c) $\Delta_{b}=2$, and then $0<\Delta_{b}-\Delta_{a}<2 (\xi_{a}+\protect\xi_{b})$.
For convenience, all the parameters are in units of $\protect\xi_{a}$
and we always set $\protect\xi_{b}=\protect\xi_{a}=1$, $\Delta_{a}=0$, $\Omega=1$, and $g_{a}=g_{b}=0.5$.}
\label{fig:2}
\end{figure}

In Fig.~\ref{fig:2},  we plotted the transmittance $T^{a}(E)\equiv |t^{a}(E)|^{2}$,
transfer rate $T^{b}\equiv \left\vert t^{b}(E)\right\vert ^{2}$ and
reflectance $R^{a}(E)\equiv |r^{a}(E)|^{2}$ as function
of the incident energy $E$. Three different band configurations are
presented. In Fig.~\ref{fig:2}(a), there is no overlap between two bands. In
this case, single photons can not travel in CRW-$b$, because $k_{b}$ is
complex with positive imaginary component, i.e., single photons are
localized around the atom to form local modes in CRW-$b$. Therefore, the photon flow is confined
in CRW-$a$, generating the flow conservation equation $\left\vert
t^{a}\right\vert ^{2}+\left\vert r^{a}\right\vert ^{2}=1$. However, single
photons in CRW-$a$ can be perfectly reflected when their energies
match the eigenvalues of the bound states of CRW-$b$ with $g_{a}=0$.
In Fig.~\ref{fig:2}(b), there is the maximum overlap between two
bands. In this case, the flow conservation relation changes into
$|t^{a}| ^{2}+\left\vert r^{a}\right\vert ^{2}+2|t^{b}|^{2}=1$.
The non-vanishing transfer coefficients $T^{b}$ and shows that the
classical field redirects the single photons coming from one continuum to
the other. In Fig.~\ref{fig:2}(c), there is partial overlap between two
bands. When the energy of the incident photon is out of (within) the overlap
region of the two continuum bands, the conservation relation and the related
scattering properties are the same as that in Fig.~\ref{fig:2}(a)
[Fig.~\ref{fig:2}(c)]. We note that the coefficient $T^{b}$
can be greater than $1$ for the bound state of CRW-$b$. It is similar to the Feshbach resonance
in cold atom scattering that the scattering cross section diverges when
the energy of the incident particle matches the bound state of the closed channel.
Here, the bound states in CRW-b form closed channel-b and $T^b$ represents the amplitude
of the bound states localized around the $\bigtriangleup$-atom instead of the
transmission coefficient.

\textit{Quantum routing for single photons}.---To demonstrate the routing functions
of our hybrid system, we first
investigate the reflectance, transmittance and transfer rate of single photons in
detail when two bands are maximally overlapped.

\begin{figure}[tbp]
\includegraphics[width=4cm]{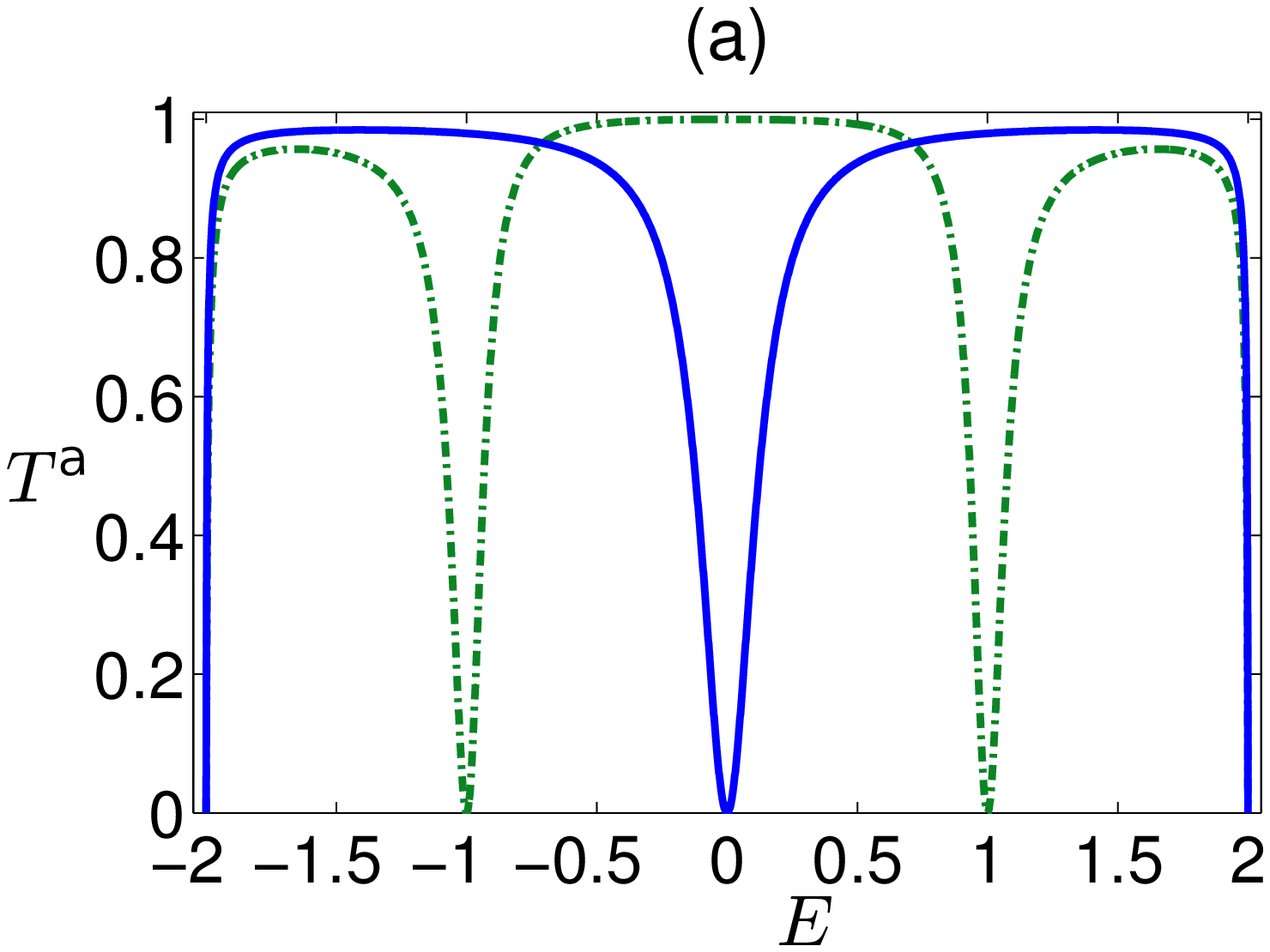}\includegraphics[width=4cm]{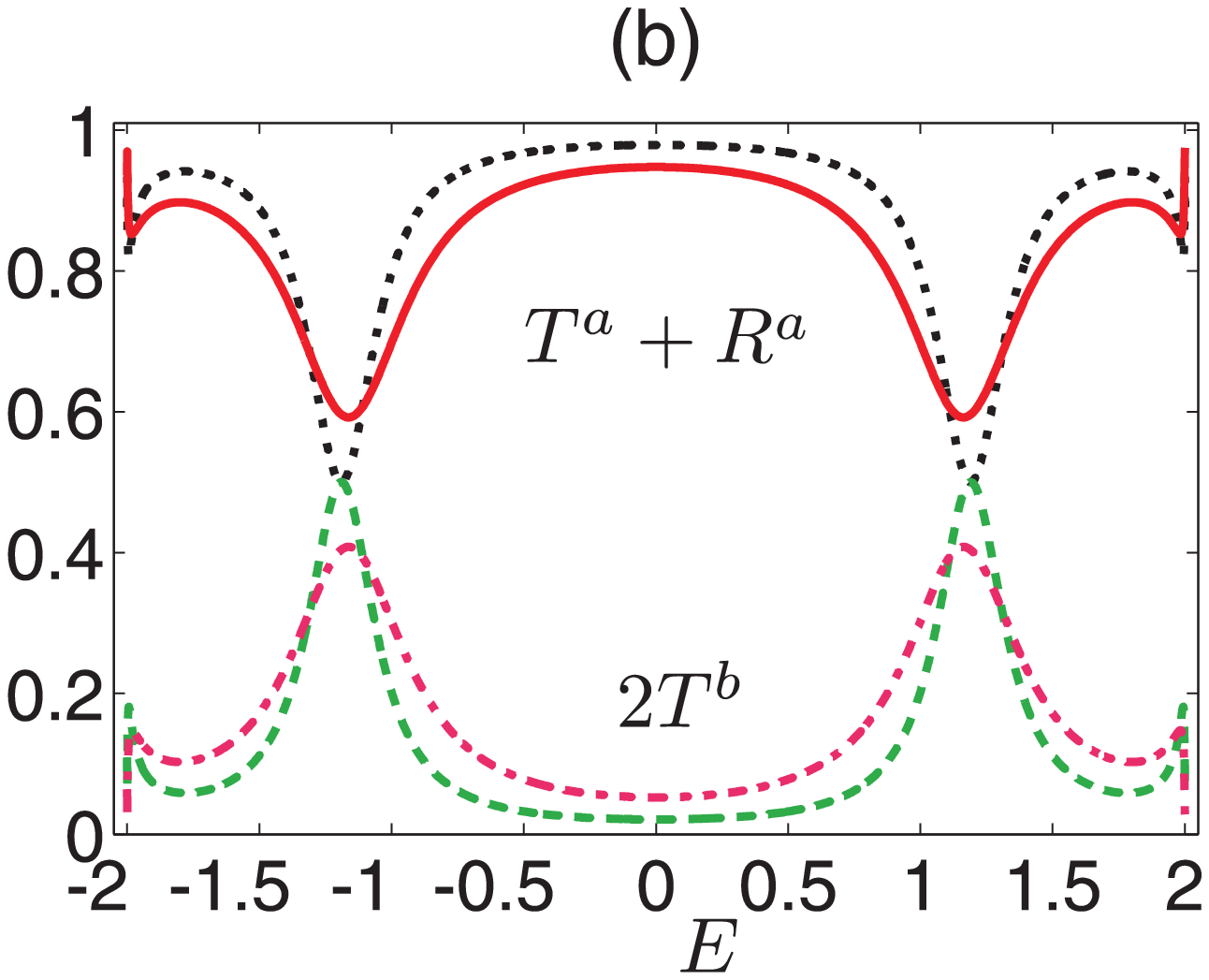} %
\includegraphics[width=4cm]{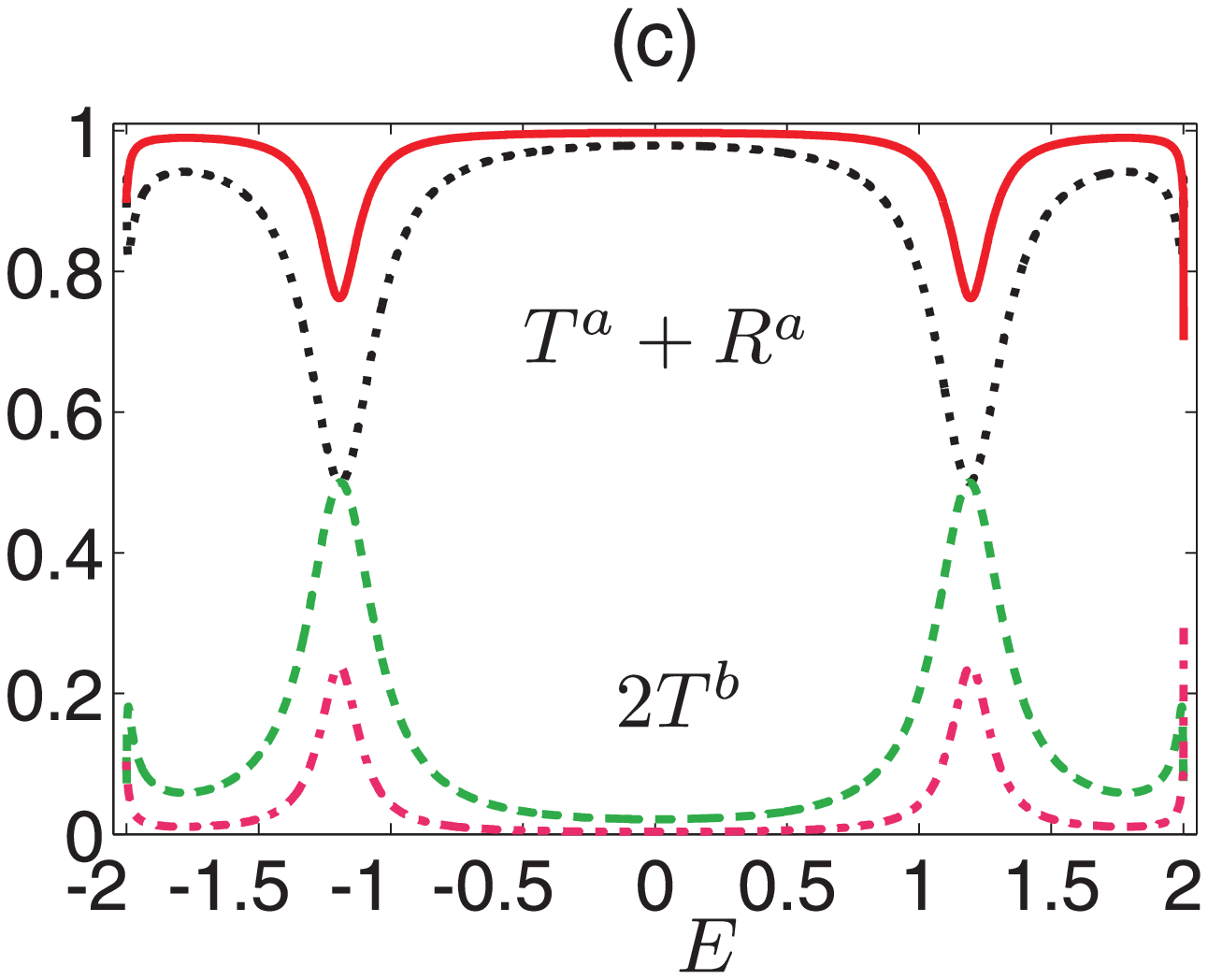}\includegraphics[width=4cm]{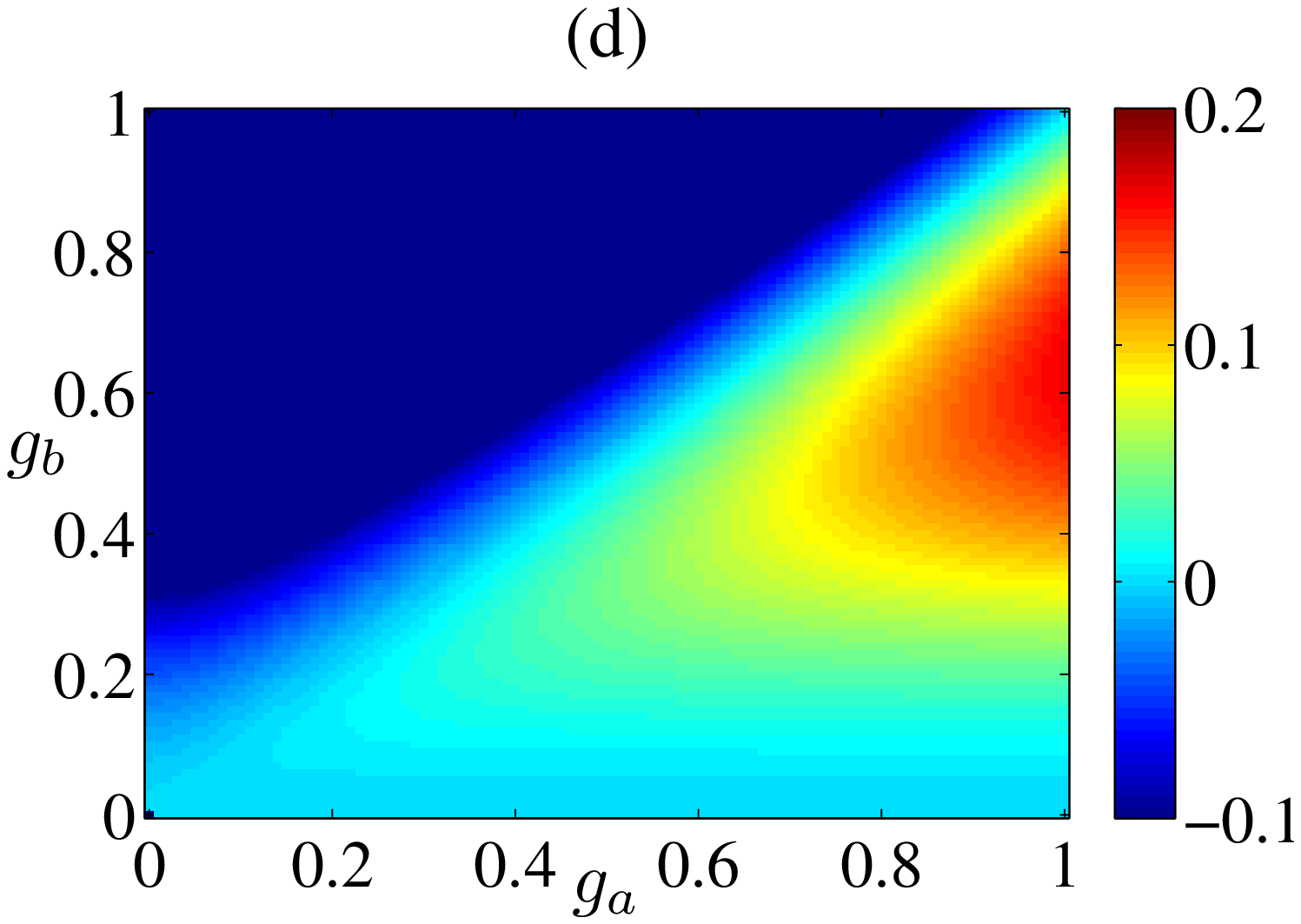}
\caption{(Color online) (a) The transmission $T^{a}$ as a function of the
incident energy $E$ with $g_{a}=0.5$ and $\Omega=0$ for solid
blue line, or with $g_{a}=0.5$, $g_{b}=0$ and $\Omega=1$ for dotted-dashed green line;
(b) [(c)] The coefficients $T^{a}+R^a$ (solid red and dotted black lines), $2T^{b}$
(dashed green and dotted-dashed pink lines) as functions of the incident energy
$E$ with $g_{a}=0.5$ and $\Omega=1.2$, $g_{b}=0.5$ for dotted black and dashed
green lines, or $g_{b}=0.8 [0.2]$ for solid red and dotted-dashed pink lines.
(d) Transmittance difference $(T^{b}-T^{a})$ at the condition $\left|E\right|=\Omega$
vs $g_{a}$ and $g_{b}$. Here, we take $\Delta_{a}=\Delta_{b}=0,$ and $\xi_{a}=\xi_{b}=1$.}
\label{fig:3}
\end{figure}

When $\Omega=0$, CRW-$a$ and -$b$ are decoupled for the case with single excitation,
thus an incoming wave is effectively scattered by a two-level atom ($\left|g\right\rangle
\leftrightarrow\left|e\right\rangle $) within the incident channel. As shown in
Fig.~\ref{fig:3}(a) (the solid blue line), the perfect reflection only occurs when
incident waves resonate with the corresponding atomic transition~\cite{sun2l}. The
green dot-dashed line in Fig.~\ref{fig:3}(a) describes the scattering process where
an incoming wave in CRW-$a$ encounters a three-level atom in the absence of CRW-$b$
(i.e., $g_{b}=0$). It is found that the classical field makes the solid blue line
split into a doublet with a separation of $2\Omega$. That means the incident photon
resonant with the  atomic transition $|e\rangle\leftrightarrow |g\rangle$, becomes
transparent. Actually, $|E|=\Omega$ correspond to two dressed states of the atom
induced by the classical driving $\Omega$. Therefore, by resonantly driving the
transition $|e\rangle\leftrightarrow |f\rangle$, we are allowed to observe the
electromagnetically induced transparency (EIT) based on the Autler-Townes splitting~\cite{ATS}.
When another output channel (CRW-b) exists, the conservation relation of the photon
flow becomes $\left|t^{a}\right|^{2}+\left|r^{a}\right|^{2}+2\left|t^{b}\right|^2=1$.
In Figs.~\ref{fig:3}(b) and (c), the sufficiently large transfer rate $T^{b}$ shows
that the classical field can redirect the single photons coming from one CRW to the
other and its strength $\Omega$ determines the position where the minimum of transmission
in CRW-$a$ and the maximum of the probability transferred to CRW-$b$ occur in the
energy axis. That is, $2T^b$ has two peaks centered at $E=\pm\Omega$. The height of these
peaks for $2T^b$ will take maximal values when $g_a=g_b$. We can further find that
the coupling strengths $g_a$ and $g_b$ also determine the width of each peaks for $2T^b$:
The larger the product $g_ag_b$ are, the wider each peaks are. The transmittance difference
$T^{b}-T^{a}$ at $\left|E\right|=\Omega$ in Fig.~\ref{fig:3}~(d) shows the influence
of the coupling strengths $g_a$ and $g_b$. Although the infinite potential
$V_{a}\left(\pm\Omega\right)$ is supposed to totally reflect the incident single photons,
the infinite coupling $G\left(\pm\Omega\right)$ makes the transmission zeros disappear.
It can be found that, when $g_{a}=g_{b}$,
$|V_{a}(\pm\Omega)|=|V_{b}(\pm\Omega)|=|G\left(\pm\Omega\right)|$ and then $T^{a}(\pm\Omega)=T^{b}(\pm\Omega)$.

\begin{figure}[tbp]
\includegraphics[width=8cm]{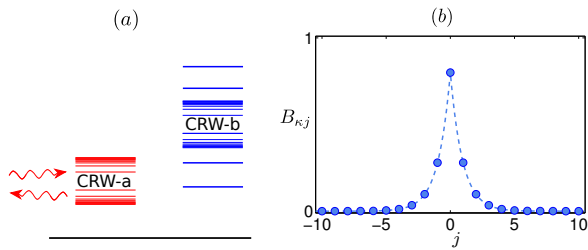}
\caption{(Color online) (a) The CRW-b plus the $\bigtriangleup$-atom have two bound states
both above and blow the continuum band. The incident photon get perfectly reflected when its energy
matches the bound states within CRW-b; (b) Wave function of bound state within CRW-b with $n_{b}=0$.}
\label{fig:4}
\end{figure}

\textit{Bound states}.---Now we come to explore the underlying physical
mechanism of perfect reflection in Fig.~\ref{fig:2}(a) and (c).
Although propagating photons can get transmitted in CRW-$a$ or
reflected by the three-level atom, the perfect reflections in Fig.~\ref{fig:2}(a)
do not appear at the same position as in Fig.~\ref{fig:3}(a).
Here, we first study the bound states of CRW-b plus the driven $\bigtriangleup$-atom (Fig.~\ref{fig:4}(a)).
The bound states exist when the translation symmetry of the CRW-$b$ is broken.
By applying the spatial-exponentially-decay solution
$B_{\kappa j}=C\exp(in_{b}\pi j-\kappa|j|)$ ($\kappa>0$ and $n_{b}=0,1$) as shown in Fig.~\ref{fig:4}(b) to
the scattering equation Eq.~(\ref{eq:scat}) with $g_{a}=0$, we obtain the
self-consistent condition for the energies of the bound states of CRW-$b$,
\begin{equation}
(-1)^{n_{b}}(E^{2}-\Omega^{2})\sqrt{(E-\Delta_{b})^{2}-4\xi_{b}^{2}}%
+Eg_{b}^{2}=0.  \label{eq:CRWB-BS}
\end{equation}
Solutions of Eq.~(\ref{eq:CRWB-BS}) with $n_{b}=0$ ($n_{b}=1$) indicates
that the eigen-energies lie below (above) the band. Thus, when the incident
energy are out of the energy band of CRW-$b$, these photons can not travel
out of this channel, resulting in the flow conservation $\left|t^{a}%
\right|^{2}+\left|r^{a}\right|^{2}=1$.

We note that the left-hand side of Eq.~(\ref{eq:CRWB-BS}) is exactly the
term in the square bracket of the numerator in Eq.~(\ref{eq:tal-to-r}) by
replacing $k_{b}$ with $n_{b}\pi+i\kappa$, i.e., the transmission zeros of
$T^{a}$ are completely determined by the bound states. Here, the applied
classical field couples the bound states of CRW-$b$ to the continuum of CRW-$a$,
then bound states plus the continuum band of CRW-a forms quasi-bound states of the
total system~\cite{Dazhi}. Two interfering paths are formed: single photons travel
directly through the CRW-$a$, or visit the bound states, return back, and continue to propagate
in CRW-$a$. When the incident photons resonate with one of the lower two bound states,
the interference originated from the interaction of a discrete localized
state with a continuum of guiding modes leads to the total reflection of
single photons (Fig.~\ref{fig:4}(a)). This is different from the cases in Fig.~\ref{fig:3} whose zeros ($T^{a}=0$) result from the
coherent interference between the incoming wave and the wave scattered from
the atom~\cite{sun2l}. Furthermore, this effect can be used to probe these
bound states.

\textit{Conclusions}.---We have studied the coherent scattering process of
single photons in two 1D CRWs by a $\bigtriangleup$-type atom, which behaves
as a quantum multi-channel router. Here, the $\bigtriangleup$-atom functions
as single-photon switch within the incident channel with the following two ways:
1) adjusting the transition energy of the artificial atom associated to the
incident channel in the absence of the classical field; 2) adjusting the
configuration of the two continuums by changing the detunings to make the
incident photons match the bound states of the other CRW in the presence of
the classical field. When the classical field is applied to dress the atom,
single photons can be routed from one channel to the other once any dressed
state matches the continuums of the two channels. The promising candidates
for experimental implementations of the above quantum routing system are the
following: The circuit QED system~\cite{CirQED} where two coplanar linear
resonators are coupled to a cyclic $\bigtriangleup$-atom~\cite{cycleAL}
using three Josephson junctions and microwaves serve as the classical
controlling field; The defect cavities in photonic crystal coupled to a
silicon-based quantum dot~\cite{2CRW,dotinres}. It is much appreciated that
the quantum routing function we predicted can be observed in some
experiments based on such hybrid systems.

This work is supported by NSFC No. 11074071, No.~11121403, No.~10935010,
No.~11174027, and No. 11074261, NFRPC No. 2012CB922103 and No. 2012CB922104,
and Hunan Provincial Natural Science Foundation of China (11JJ7001,
12JJ1002).

\end{document}